\documentclass[12pt]{article}
\usepackage{a4wide}
\usepackage{epsfig}
\usepackage{amsmath}

\newlength{\absize}
\catcode`@=11
\def\citer{\@ifnextchar [{\@tempswatrue\@citexr}{\@tempswafalse\@citexr[]}}

\def\@citexr[#1]#2{\if@filesw\immediate
  \write\@auxout{\string\citation{#2}}\fi
  \def\@citea{}\@cite{\@for\@citeb:=#2\do
    {\@citea\def\@citea{--\penalty\@m}\@ifundefined
       {b@\@citeb}{{\bf ?}\@warning
       {Citation `\@citeb' on page \thepage \space undefined}}%
\hbox{\csname b@\@citeb\endcsname}}}{#1}}
\catcode`@=12

\begin{document}
  \thispagestyle{empty}
  \pagestyle{empty}
  \renewcommand{\thefootnote}{\fnsymbol{footnote}}
\newpage\normalsize
    \pagestyle{plain}
    \setlength{\baselineskip}{4ex}\par
    \setcounter{footnote}{0}
    \renewcommand{\thefootnote}{\arabic{footnote}}
\newcommand{\preprint}[1]{%
  \begin{flushright}
    \setlength{\baselineskip}{3ex} #1
  \end{flushright}}
\renewcommand{\title}[1]{%
  \begin{center}
    \LARGE #1
  \end{center}\par}
\renewcommand{\author}[1]{%
  \vspace{2ex}
  {\Large
   \begin{center}
     \setlength{\baselineskip}{3ex} #1 \par
   \end{center}}}
\renewcommand{\thanks}[1]{\footnote{#1}}
\begin{flushright}
\end{flushright}
\vskip 0.5cm

\begin{center}
{\large \bf Un-equivalency Theorem between Deformed and undeformed
Heisenberg-Weyl's Algebras}
\end{center}
\vspace{1cm}
\begin{center}
Jian-Zu Zhang
\end{center}
\vspace{1cm}
\begin{center}
Institute for Theoretical Physics, East China University of
Science and Technology, \\
Box 316, Shanghai 200237, P. R. China
\end{center}
\vspace{1cm}
\begin{abstract}
Two fundamental issues about the relation between the deformed
Heisenberg-Weyl algebra in noncommutative space and the undeformed
one in commutative space are elucidated. First the un-equivalency
theorem between two algebras is proved: the deformed algebra
related to the undeformed one by a non-orthogonal similarity
transformation is explored; furthermore, non-existence of a
unitary similarity transformation which transforms the deformed
algebra to the undeformed one is demonstrated. Secondly the
uniqueness of realizing the deformed phase space variables via the
undeformed ones is elucidated: both the deformed Heisenberg-Weyl
algebra and the deformed bosonic algebra should be maintained
under a linear transformation between two sets of phase space
variables which fixes that such a linear transformation is unique.
Elucidation of this un-equivalency theorem has basic meaning both
in theory and experiment.
\end{abstract}


\clearpage

Spatial noncommutativity is an attractive basic idea for a long
time. Recent interest on this subject is motivated by studies of
the low energy effective theory of D-brane with a nonzero NS - NS
$B$ field background \citer{DH,DN}. It shows that such low energy
effective theory lives on noncommutative space. For understanding
low energy phenomenological events quantum mechanics in
noncommutative space (NCQM) is an appropriate framework. NCQM have
been extensively studied and applied to broad fields
\citer{CST,WLFXY}. But up to now it is not fully understood.

In literature there is an extensively tacit understanding about
equivalency between the deformed Heisenberg-Weyl algebra in
noncommutative space and the undeformed one in commutative space.
As is well known, the deformed phase space variables are related
to the undeformed ones by a linear transformation, thus one
concludes that the algebra of noncommutative quantum mechanical
observables is the standard one. This leads to the tacit
understanding of fully equivalency between two algebras. A related
tacit understanding is that there are many equivalent linear
transformations between two sets of phase space variables.

In this paper we elucidate these two subtle points. First we
clarify equivalency conditions between two algebras. We
demonstrate that the deformed algebra is related to the undeformed
one by a similarity transformation with a non-orthogonal real
matrix. Furthermore, we prove that a unitary similarity
transformation which transforms two algebras to each other does
not exist. The results are summarized in the un-equivalency
theorem between two algebras. Secondly we clarify that 
among deferent types of linear transformations of realizing
deformed phase space variables via undeformed ones only a unique
one maintains both the deformed Heisenberg-Weyl algebra and the
deformed bosonic algebra.

In order to develop the NCQM formulation we need to specify the
phase space and the Hilbert space on which operators act. The
Hilbert space is consistently taken to be exactly the same as the
Hilbert space of the corresponding commutative system \citer{CST}.
As for the phase space we consider both position-position
noncommutativity (position-time noncommutativity is not
considered) and momentum-momentum noncommutativity
\cite{DN,JZZ04a}. In this case the consistent deformed
Heisenberg-Weyl algebra is as follows:
\begin{equation}
\label{Eq:xp}
[\hat x_{i},\hat x_{j}]=i\xi^2\theta\epsilon_{ij},
\qquad [\hat x_{i},\hat p_{j}]=i\hbar\delta_{ij}, \qquad
[\hat p_{i},\hat p_{j}]=i\xi^2\eta\epsilon_{ij},\;(i,j=1,2),
\end{equation}
where $\theta$ and $\eta$  are the constant, frame-independent
parameters.
Here we consider the intrinsic momentum-momentum noncommutativity.
It means that the parameter $\eta$, like the parameter $\theta$,
should be extremely small.
$\epsilon_{ij}$ is an antisymmetric unit tensor,
$\epsilon_{12}=-\epsilon_{21}=1,$ $\epsilon_{11}=\epsilon_{22}=0$;
$\xi=(1+\theta\eta/4\hbar^2)^{-1/2}$ is the scaling factor.
For the case of both position - position and momentum - momentum
noncommuting the scaling factor $\xi$ in Eq.~(\ref{Eq:xp})
guarantees consistency of the framework, and plays an essential
role in dynamics as well. For example, in the discussion of
deformed two - mode quadrature operators it revealed that effects
of spatial noncommutativity are included in the scaling factor
$\xi$ \cite{WLFXY}.
When $\eta=0,$ we have $\xi=1$. The deformed Heisenberg-Weyl
algebra (\ref{Eq:xp}) reduces to the one of only position-position
noncommuting.

The deformed phase space variables $\hat x_{i}$ and $\hat p_{i}$
are related to the undeformed ones $x_{i}$ and $p_{i}$ by the
following linear transformation \cite{JZZ04a}
\begin{equation}
\label{Eq:hat-x-p}
\hat x_{i}=\xi(x_{i}-\frac{1}{2\hbar}\theta\epsilon_{ij}p_{j}),
\quad
\hat p_{i}=\xi(p_{i}+\frac{1}{2\hbar}\eta\epsilon_{ij}x_{j}).
\end{equation}
where $x_{i}$ and $p_{i}$ satisfy the undeformed Heisenberg-Weyl
algebra
$[x_{i},x_{j}]=[p_{i},p_{j}]=0, [x_{i},p_{j}]=i\hbar\delta_{ij}.$

In literature the point of the tacit understanding of equivalency
between the deformed Heisenberg-Weyl algebra and the undeformed
one is as follows: any Lie algebra generated by relations
$[X_a,X_b]=iT_{ab}$ with central $T_{ab}$ satisfying
$det(T_{ab})\ne 0$ can be put into a usual canonical form, like
Eqs.~(\ref{Eq:hat-x-p}). Therefore the deformed Heisenberg-Weyl
algebra (\ref{Eq:xp}) and the undeformed one
are
the same, and the spectrum of an observable is the same regardless
we star with deformed variables $(\hat x_{i},\hat p_{i})$ or
undeformed ones $(x_{i},p_{i})$.

Now we elucidate this subtle point. Equivalency between the
deformed Heisenberg-Weyl algebra and the undeformed one must
satisfy two conditions: (i) Two sets of phase space variables
$(\hat x_{i}, \hat p_{i})$ and $(x_{i}, p_{i})$ can be related to
each other by a singular-free linear transformation (The inverse
transformation should exit for all values of $(\hat x_{i}, \hat
p_{i})$ and $(x_{i}, p_{i})$);
(ii) Two algebras can be transformed to each other by a unitary
similarity transformation.

First we consider the second condition. We prove the following
theorem.

{\bf The Un-equivalency Theorem} {\it The deformed Heisenberg-Weyl
algebra in noncommutative space is transformed to the undeformed
one in commutative space by a similarity transformation with a
non-orthogonal real matrix. A unitary similarity transformation
which relates two algebras to each other does not exist.}

The demonstration of the first part of the theorem is trivial. We
define a $1\times 4$ column matrix $\hat U=(\hat U_1,\hat U_2,\hat
U_3,\hat U_4)$ with elements $\hat U_1=\hat x_1$, $\hat U_2=\hat
x_2$, $\hat U_3=\hat p_1$ and $\hat U_4=\hat p_2$, a $4\times 1$
row matrix $\hat U^T$ with elements $\hat U_i^T=\hat U_i$,
$(i=1,2,3,4)$, and a $4\times 4$ matrix $\hat M$ with elements
$i\hat M_{ij}=[\hat U_i, \hat U_j^T], (i, j=1,2,3,4).$
The matrix $\hat M$ represents the deformed Heisenberg-Weyl
algebra. From Eqs.~(\ref{Eq:xp})
it follows
that $\hat M$ reads
\begin{equation}
\label{Eq:hat-M} 
\mathbf{\hat M}=\left(\begin{array}{cccc}
0&\xi^2\theta&\hbar&0 \\
-\xi^2\theta&0&0&\hbar \\
-\hbar&0&0&\xi^2\eta \\
0&-\hbar&-\xi^2\eta&0
\end{array}\right).
\end{equation}
The corresponding matrixes in commutative space are a $1\times 4$
column matrix $U$ with elements $U_1=x_1$, $U_2=x_2$, $U_3=p_1$
and $U_4=p_2$, a $4\times 1$ row matrix $U^T$ with elements
$U_i^T=U_i$, $(i=1,2,3,4)$, and a $4\times 4$ matrix $M$ with
elements
$iM_{ij}=[U_i, U_j^T], (i, j=1,2,3,4).$
The matrix $M$ represents the undeformed Heisenberg-Weyl algebra,
which can be obtained by putting $\theta=\eta=0$ in the matrix
$\hat M$ (\ref{Eq:hat-M}),
\begin{equation}
\mathbf{M}=\left(\begin{array}{cccc}
0&0&\hbar&0 \\
0&0&0&\hbar \\
-\hbar&0&0&0 \\
0&-\hbar&0&0  \nonumber
\end{array}\right).
\end{equation}

From Eq.~(\ref{Eq:hat-x-p}) it follows that $\hat U_i=R_{ik}U_k$,
$\hat U^T_j=\hat U_j==R_{jl}U_l=U^T_lR^T_{lj}$, and the deformed
Heisenberg-Weyl algebra is related to the undeformed
Heisenberg-Weyl algebra by a similarity transformation
$\hat M_{ij}=R_{ik}M_{kl}R^T_{lj}$
 with a real matrix $R$
\begin{equation}
\label{Eq:matrix-R} 
\mathbf{R}=\left(\begin{array}{cccc} \xi&
0&0&-\frac{1}{2\hbar}\xi\theta \\
0&\xi&\frac{1}{2\hbar}\xi\theta&0 \\
0&\frac{1}{2\hbar}\xi\eta&\xi&0 \\
-\frac{1}{2\hbar}\xi\eta&0&0&\xi
\end{array}\right).
\end{equation}
It is obvious that $R$ is not orthogonal matrix $RR^T\ne I$.

Now we prove the second part of the un-equivalency theorem.
Eq.~(\ref{Eq:hat-x-p}) shows that if there is such a unitary
transformation, its elements should be real. That is, it should be
an orthogonal matrix $S$ with real elements $S_{ij}$,
$SS^T=S^TS=I$, and satisfies $S_{ik}\hat M_{kl} S^T_{lj}=M_{ij}$,
or $S_{ik}\hat M_{kj}=M_{ik} S_{kj}$. This is a system of 16
homogeneous linear equations for $S_{ij}$, $(i,j=1, 2, 3, 4)$. It
is divided into 4 closed sub-systems of 4 homogeneous linear
equations. Among them we consider a closed sub-system including
$S_{12}$, $S_{13}$, $S_{31}$ and $S_{34}$, which reads
\begin{subequations}
\begin{equation}
\label{Eq:4eqs-a}
\xi^2\theta S_{12}+\hbar S_{13}=-\hbar S_{31},\,
\end{equation}
\begin{equation}
\label{Eq:4eqs-b}%
\hbar S_{12}+\xi^2\eta S_{13}=\hbar S_{34},\,
\end{equation}
\begin{equation}
\label{Eq:4eqs-c}%
\xi^2\theta S_{31}-\hbar S_{34}=-\hbar S_{12},\,
\end{equation}
\begin{equation}
\label{Eq:4eqs-d}%
\hbar S_{31}-\xi^2\eta S_{34}=-\hbar S_{13}.
\end{equation}
\end{subequations}
The condition of non-zero solutions of $S_{12}$, $S_{13}$,
$S_{31}$ and $S_{34}$ is
\footnote {\; In Eq.~(\ref{Eq:non-zero}) dimensions of different
terms are different. If we define a $1\times 4$ column matrix
$\hat V$ with elements $\hat V_1=\hat x_1$, $\hat V_2=\hat x_2$,
$\hat V_3=\alpha\hat p_1$ and $\hat V_4=\alpha\hat p_2$, where
$\alpha$ is an auxiliary arbitrary non-zero constant with the
dimension $[mass]^{-1}[time]^1$. Thus $\hat V_i$ $(i=1, 2, 3, 4)$
have the same dimension $[length]^2$. Then in
Eq.~(\ref{Eq:non-zero}) dimensions of different terms are same.
The introduction of the arbitrary constant $\alpha$ does not
change the following conclusion.}
\begin{equation}
\label{Eq:non-zero} 
\xi^2\theta\eta=\pm \hbar(\theta+
\eta).
\end{equation}

In order to elucidate the physical meaning of
Eq.~(\ref{Eq:non-zero}), we consider conditions of guaranteeing
Bose-Einstein statistics in the case of both position-position and
momentum-momentum noncommuting in the context of non-relativistic
quantum mechanics. We start from the general construction of
deformed annihilation and creation operators $\hat a_i$ and $\hat
a_i^\dagger$ $(i=1,2)$ at the deformed level, which are related to
the deformed phase space variables $\hat x_i$ and $\hat p_i$. The
general form of $\hat a_i$ can be represented as $\hat
a_i=c_1(\hat x_i +ic_2\hat p_i)$, where constants $c_1$ and $c_2$
can be fixed as follows. The deformed annihilation and creation
operators $\hat a_i$ and $\hat a_i^\dagger$ should satisfy $[\hat
a_1,\hat a_1^\dagger]=[\hat a_2,\hat a_2^\dagger]=1$ (to keep the
physical meaning of $\hat a_i$ and $\hat a_i^\dagger$). From this
requirement and the deformed Heisenberg-Weyl algebra (\ref{Eq:xp})
it follows that $c_1=\sqrt{1/2c_2\hbar}$. When the state vector
space of identical bosons is constructed by generalizing
one-particle quantum mechanics, Bose-Einstein statistics should be
maintained at the deformed level described by $\hat a_i$, thus
operators $\hat a_1$ and $\hat a_2$ should be commuting. From
$[\hat a_i,\hat a_j]=0$ and the deformed Heisenberg-Weyl algebra
(\ref{Eq:xp}) it follows that
$ic_1^2\xi^2\epsilon_{ij}(\theta-c_2^2\eta)=0$, i.e.
$c_2=\sqrt{\theta/\eta}$. (The phases of $\theta$ and $\eta$ are
chosen so that $\theta/\eta>0$.) The general representations of
the deformed annihilation and creation operators $\hat a_i$ and
$\hat
a_i^\dagger$ are 
\begin{eqnarray}
\label{Eq:hat-a}
\hat a_i=\sqrt{\frac{1}{2\hbar}\sqrt{\frac{\eta}{\theta}}}\left
(\hat x_i +i\sqrt{\frac{\theta}{\eta}}\hat p_i\right),
\hat
a_i^\dagger=\sqrt{\frac{1}{2\hbar}\sqrt{\frac{\eta}{\theta}}}\left
(\hat x_i-i\sqrt{\frac{\theta}{\eta}}\hat p_i\right).
\end{eqnarray}
The structure of the deformed annihilation operator $\hat a_{i}$
in Eq.~(\ref{Eq:hat-a}) is determined by the deformed
Heisenberg-Weyl algebra (\ref{Eq:xp}) itself, independent of
dynamics. The special feature of a dynamical system is encoded in
the dependence of the factor $\eta/\theta$ on characteristic
parameters of the system under study.

In the limits $\theta,\eta\to 0$ and $\eta/\theta$ keeping finite,
the deformed annihilation operator $\hat a_i$ should reduce to the
undeformed annihilation operator $a_i$.
In commutative space in the context of non-relativistic quantum
mechanics the general form of the undeformed annihilation operator
$a_i$ can be represented as $a_i=d_1(x_i +id_2p_i)$. From $[a_1,
a_1^\dagger]=[a_2, a_2^\dagger]=1$ and the undeformed
Heisenberg-Weyl algebra
it also follows that $d_1=\sqrt{1/2d_2\hbar}$ with $d_2>0$. But
from the undeformed Heisenberg-Weyl algebra
the equation $[a_i,a_j]=0$ is
automatically satisfied, thus there is not constraint on the
coefficient $d_2$. The general form of the undeformed annihilation
operator reads
\begin{equation}
\label{Eq:a} 
a_i=\sqrt{\frac{1}{2d_2\hbar}}\left (x_i +id_2 p_i\right).
\end{equation}
Like the situation of the deformed annihilation operator $\hat
a_{i}$, here the structure of $a_{i}$ is determined by the
undeformed Heisenberg-Weyl algebra
itself,
independent of dynamics. The special feature of a dynamical system
is encoded in the dependence of the factor $d_2$ on characteristic
parameters of the system under study.
If noncommutative quantum theory is a realistic physics, all
quantum phenomena should be reformulated at the deformed level.
This means that in the limits $\theta,\eta\to 0$ and $\eta/\theta$
keeping finite the deformed annihilation operator $\hat a_{i}$
should reduce to the undeformed one $a_{i}$. Comparing
Eq.~(\ref{Eq:hat-a}) and (\ref{Eq:a}), it follows that in the
limits $\theta,\eta\to 0$ and $\eta/\theta$ keeping finite the
factor $\eta/\theta$ reduces to a {\it positive} quantity:
\begin{equation}
\label{Eq:eta-theta}
\frac{\eta}{\theta}\to \frac{1}{d_2^2}>0.
\end{equation}
But from Eq.~(\ref{Eq:non-zero}), we obtain
$\eta/\theta=\pm\hbar/(\xi^2\theta\mp\hbar)$. This equation shows
that in the limits $\theta,\eta\to 0$ and $\eta/\theta$ keeping
finite, we have $\eta/\theta\to -1$, which contradicts
Eq.~(\ref{Eq:eta-theta}). We conclude that Eq.~(\ref{Eq:non-zero})
is un-physical. The situation for the rest elements of $S_{ij}$ is
the same. Thus the supposed orthogonal real matrix $S$ consistent
with physical requirements does {\it not} exist. The second part
of the un-equivalency theorem is proved.

Now we consider the first condition about equivalency of the two
algebras. Eq.~(\ref{Eq:hat-x-p}) shows that the determinant
$\mathcal{R}$ of the transformation matrix $R$ between $(\hat
x_1,\hat x_2,\hat p_1,\hat p_2)$ and $(x_1,x_2,p_1,p_2)$ is
$\mathcal{R}=\xi^4(1-\theta\eta/4\hbar^2)^2.$
When $\theta\eta=4\hbar^2$, the matrix $R$ is singular. In this
case the inverse of $R$ does not exit. It means that the first
condition about equivalency of two algebras is not satisfied.

The above results show that for the case of both position-position
and momentum-momentum noncommuting the deformed Heisenberg-Weyl
algebra and the undeformed one are not equivalent.

For the case of only position-position noncommuting, $\eta=0$, the
transformation matrix $R$ between $(\hat x_1,\hat x_2,\hat
p_1,\hat p_2)$ and $(x_1,x_2,p_1,p_2)$ reduces to the matrix
\begin{equation}
\label{Eq:matrix-R0} 
\mathbf{R_0}=\left(\begin{array}{cccc} 1&
0&0&-\frac{1}{2\hbar}\theta \\
0&1&\frac{1}{2\hbar}\theta&0 \\
0&0&1&0 \\
0&0&0&1
\end{array}\right).
\nonumber
\end{equation}
Its determinant $\mathcal{R}_0\equiv 1$, which is singular-free.
But in this case $R_0$ is not an orthogonal matrix either,
$R_0R_0^T\ne I$. Furthermore, in this case the supposed orthogonal
real matrix $S$ reduces to $S_0$, which is obtained from $S$ by
setting  $\eta=0$. The closed sub-system of 4 homogeneous linear
equations including $S_{0,12}$, $S_{0,13}$, $S_{0,31}$ and
$S_{0,34}$ has only zero solutions. The supposed orthogonal real
matrix $S_0$ does {\it not} exist, either. We conclude that for
the case of only position-position noncommuting the deformed and
the undeformed Heisenberg-Weyl algebras are also not equivalent.

Now we elucidate the uniqueness of the linear realization of the
deformed phase space variables via the undeformed ones. A physical
realization should maintain both the deformed Heisenberg-Weyl
algebra and the deformed bosonic algebra.

It worth noting that among deferent types of linear
transformations between two sets of phase space variables only
Eq.~(\ref{Eq:hat-x-p}) maintains both the deformed Heisenberg-Weyl
algebra and the deformed bosonic algebra. It is trivial to check
that Eq.~(\ref{Eq:hat-x-p}) maintains the deformed Heisenberg-Weyl
algebra (\ref{Eq:xp}).

Inserting Eqs.~(\ref{Eq:hat-x-p}) into Eqs.~(\ref{Eq:hat-a}), and
using Eq.~(\ref{Eq:a}),
we obtain the linear representation of the deformed annihilation
operator by the undeformed one
\begin{equation}
\label{Eq:hat-a-a}
\hat a_{i}=\xi\left(a_{i}+\frac{i}{2\hbar}\sqrt{\theta\eta}
\epsilon_{ij}a_j\right),\;
\hat
a_{i}^\dagger=\xi\left(a_{i}^\dagger-\frac{i}{2\hbar}\sqrt{\theta\eta}
\epsilon_{ij}a_j^\dagger\right).
\end{equation}
Eq.~(\ref{Eq:hat-a-a}) maintains the deformed bosonic algebra,
including the bosonic commutation relations
$[\hat a_1,\hat a_1^\dagger]=[\hat a_2,\hat a_2^\dagger]=1.$

In literature there are another types of linear transformations
between two sets of phase space variables. One example is to set
$\xi=1$. In this case the deformed Heisenberg-Weyl algebra reduces
to:
\begin{equation}
\label{Eq:xp1} 
[\hat x_{i},\hat x_{j}]=i\theta\epsilon_{ij}, \qquad [\hat
p_{i},\hat p_{j}]=i\eta\epsilon_{ij}, \qquad
[\hat x_{i},\hat p_{j}]=i\hbar\delta_{ij},\;(i,j=1,2)
\end{equation}
and representations of deformed variables $\hat x_{i}$ and $\hat
p_{i}$ by undeformed variables $x_{i}$ and $p_{i}$ reads:
\begin{equation}
\label{Eq:hat-x-p1}
\hat x_{i}=x_{i}-\frac{1}{2\hbar}\theta\epsilon_{ij}p_{j}, \quad
\hat p_{i}=p_{i}+\frac{1}{2\hbar}\eta\epsilon_{ij}x_{j}.
\end{equation}
Inserting Eqs.~(\ref{Eq:hat-x-p1}) into Eqs.~(\ref{Eq:xp1}), the
Heisenberg commutation relation in Eqs.~(\ref{Eq:xp1}) is changed
to
\begin{equation}
\label{Eq:xp2} 
[\hat x_{i},\hat p_{j}]
=i\hbar\left(1+\frac{\theta\eta}{4\hbar^2}\right)\delta_{ij}.
\end{equation}
In order to maintain the Heisenberg commutation relation,
one may introduces
an effective Planck constant
$\hbar_{eff}=\hbar\left(1+\theta\eta/4\hbar^2\right)$
and explains $\hbar_{eff}$ as a modification of the Planck
constant by spatial noncommutativity. In order to clarify the real
physical meaning of Eq.~(\ref{Eq:xp2}) we consider the linear
representation of the deformed annihilation operator by the
undeformed one again. By the similar procedure of leading to
Eq.~(\ref{Eq:hat-a}), for the case $\xi=1$ we obtain
\begin{equation}
\label{Eq:hat-a-a1} 
\hat a_{i}=a_{i}+\frac{i}{2\hbar}\sqrt{\theta\eta}
\epsilon_{ij}a_j,\quad
\hat a_{i}^\dagger=a_{i}^\dagger-\frac{i}{2\hbar}\sqrt{\theta\eta}
\epsilon_{ij}a_j^\dagger.
\end{equation}
Eq.~(\ref{Eq:hat-a-a1}) leads to the following bosonic commutation
relations
\begin{equation}
\label{Eq:hat-a-a2} 
[\hat a_1,\hat a_1^\dagger]=[\hat a_2,\hat a_2^\dagger]
=(1+\frac{\theta\eta}{4\hbar^2}).
\end{equation}
Eq.~(\ref{Eq:hat-a-a1}) does not maintain the bosonic commutation
relations
$[\hat a_1,\hat a_1^\dagger]=[\hat a_2,\hat a_2^\dagger] =1.$
The physical meaning of Eq.~(\ref{Eq:xp2}) is similar to
Eq.~(\ref{Eq:hat-a-a2}). The correct physical explanation of
Eq.~(\ref{Eq:xp2}) is that Eq.~(\ref{Eq:hat-x-p1}) does not
maintain the Heisenberg commutation relation $[\hat x_{i},\hat
p_{j}]=i\hbar\delta_{ij}.$

We can demonstrate that except Eq.~(\ref{Eq:hat-x-p}) any other
type of linear transformations between two sets of phase space
variables can't maintain both the deformed Heisenberg-Weyl algebra
and the deformed bosonic algebra.


Because the deformed Heisenberg-Weyl algebra and the undeformed
one are, respectively, the foundations of noncommutative quantum
theories and commutative ones, elucidation of the un-equivalency
between two algebras has significant meaning both in theories and
experiments. Based on such a un-equivalency one can expect
essentially new effects of spatial noncommutativity emerged from
noncommutative quantum theories.

\vspace{0.4cm}

This work has been supported by the Natural Science Foundation of
China under the grant number 10575037 and by the Shanghai
Education Development Foundation.

\clearpage


\end{document}